\documentclass[12pt]{article}
\usepackage{fullpage}
\usepackage[authoryear]{natbib}
\usepackage[utf8]{inputenc}
\usepackage{amsmath}
\usepackage{amssymb}
\usepackage{amsfonts,}
\usepackage{bbm}
\usepackage{graphicx}
\usepackage{float}
\usepackage{caption}
\usepackage{subcaption,mathrsfs,multirow}
\usepackage[ruled]{algorithm2e}
\usepackage{array}
\usepackage[title]{appendix}
\usepackage{colortbl}

\usepackage{xr}
\usepackage{verbatim}
\usepackage{listings}
\usepackage{color}
\usepackage{epsfig}
\usepackage{wasysym}

\usepackage{color}

\newcommand{\comout}[1]{}

\def\subopt{_{\mbox{\tiny opt}}}
\def\gopt{g\subopt}

\def\Lsc{\mathcal{L}}

\def\omegahat{\widehat{\omega}}
\def\Ghat{\widehat{G}}

\def\lpbox#1{\vskip1mm \begin{center}
        \hspace{.0\textwidth}\vbox{\hrule\hbox{\vrule\kern6pt
\parbox{.95\textwidth}{\kern6pt \blue #1 (LP)\vskip6pt}\kern6pt\vrule}\hrule}
        \end{center} \vskip-5mm}

\usepackage[usenames,dvipsnames,svgnames,table]{xcolor}
\usepackage[colorlinks=true,
            linkcolor=red,
            urlcolor=blue,
            citecolor=blue]{hyperref}

\graphicspath{{./}{./Figures/}}

\usepackage{setspace}

\definecolor{lbcolor}{rgb}{0.95,0.95,0.95}
\def\EE{\mathbb{E}}

\def\half{\frac{1}{2}}

\def\Gbb{\mathbb{G}}

\def\PTE{\mbox{PTE}}
\def\Low{\mbox{Low}}
\def\PTEhat{\widehat{\mbox{PTE}}}

\def\ghat{\widehat{g}}

\def\muhat{\widehat{\mu}}

\def\Deltahat{\widehat{\Delta}}
\def\nhalf{n^{1/2}}

\def\supone{^{(1)}}
\def\supzero{^{(0)}}
\def\supa{^{(a)}}
\def\Lsc{\mathcal{L}}

\def\nhalf{n^{\half}}

\def\bD{\mathbf{D}}

\definecolor{darkred}{RGB}{150,50,50}
\definecolor{brown}{RGB}{250,100,100}
\definecolor{green}{RGB}{000,150,100}
\definecolor{purple}{RGB}{200,000,250}

\def\blue{\color{blue}}

\def\blue{\color{blue}}

\begin{document}

\title{Note}
\begin{center}
{\large \bf Model-free Approach to Evaluate a Censored Intermediate Outcome as a Surrogate for Overall Survival} \vspace{.2in}

Xuan Wang \\
{\em Division of Biostatistics, Department of Population Health Sciences, University of Utah, Salt Lake City, UT 84108, USA} \\[2ex]

Tianxi Cai \\
{\em Department of Biostatistics, Harvard University,  Boston, MA 02115, USA }\\[2ex]
 
{Lu Tian} \\
{\em Department of Biomedical Data Science, Stanford University, Stanford, CA 94305, USA}  \\[2ex]

Layla Parast \\
{\em Department of Statistics and Data Science, University of Texas at Austin, Austin, TX 78712, USA} \\
Correspondence: parast@austin.utexas.edu
\end{center}

\begin{abstract}
Clinical trials or studies oftentimes require long-term and/or costly follow-up of participants to evaluate a novel treatment/drug/vaccine. There has been increasing interest in the past few decades in using short-term surrogate outcomes as a replacement of the primary outcome i.e., in using the surrogate outcome, which can potentially be observed sooner, to make inference about the treatment effect on the long-term primary outcome.  Very few of the available statistical methods to evaluate a surrogate are applicable to settings where \textit{both} the surrogate and the primary outcome are time-to-event outcomes subject to censoring. Methods that can handle this setting tend to require parametric assumptions or be limited to assessing only the restricted mean survival time. In this paper, we propose a non-parametric approach to evaluate a censored surrogate outcome, such as time to progression, when the primary outcome is also a censored time-to-event outcome, such as time to death, and the treatment effect of interest is the difference in overall survival. Specifically, we define the proportion of the treatment effect on the primary outcome that is explained (PTE) by the censored surrogate outcome in this context, and estimate this proportion by defining and deriving an optimal transformation of the surrogate information. Our approach provides the added advantage of relaxed assumptions to guarantee that the true PTE is within (0,1), along with being model-free. Finite sample performance of our estimators are illustrated via extensive simulation studies and a real data application examining progression-free survival as a surrogate for overall survival for patients with metastatic colorectal cancer.
 
\ \
 
\end{abstract}

\newpage
\clearpage

\setstretch{1.5}

\section{Introduction}
Randomized clinical trials (RCTs) are often considered the gold standard in evaluating the effectiveness of a new treatment or drug compared to the standard care or placebo. In studies of chronic diseases, the primary outcome is typically the time to the occurrence of a clinical event and usually requires long-term follow-up. Such follow-up, while necessary, increases study costs, duration, and patient burden. Thus, unsurprisingly, there has been increasing interest in the past few decades in using short-term surrogate outcomes as a replacement of the primary outcome. That is, in using the surrogate outcome, which can potentially be observed sooner, to make inference about the treatment effect on the long-term primary outcome. To be sure, for better or for worse, short-term surrogate outcomes are currently used in some clinical trials, with approval by the Food \& Drug Administration (FDA) in the US \citep{FDA_approved}. For example, in diabetes studies, reaching elevated levels of hemoglobin A1c or fasting plasma glucose  have been used as surrogates when the primary outcome is time to a diabetes diagnosis \citep{us2008guidance}. In addition, a diagnosis of pre-diabetes is considered a potential surrogate for diabetes \citep{alberti2006metabolic, lorenzo2003metabolic}. In cardiovascular research, non-fatal cardiovascular outcomes, such as a myocardial infarction, stroke, or congestive heart failure, have also been considered as surrogates for overall death \citep{allhat2000major, ridker2005randomized}.

Certainly, there is immense risk in using a surrogate outcome to make a decision about the effectiveness of a treatment. One could mistakenly conclude there is a treatment effect on the primary outcome based on surrogate outcome information only, when in fact there is no treatment effect on the primary outcome; or one could conclude there is no treatment effect when there truly is. The most dangerous situation is if one concludes there is a positive treatment effect on the primary outcome, when in fact, there is a negative, perhaps deadly, effect on the primary outcome. Such errors have huge potential implications in terms of lives and costs, and highlight the importance of rigorous statistical methods to evaluate a surrogate outcome.

 Fortunately, many useful measures/methods have been proposed to evaluate the surrogacy of a marker or outcome being considered as a potential surrogate. These include estimation of indirect and direct effects \citep{robins1992identifiability}, average causal necessity, average causal sufficiency and the causal effect predictiveness in a principal stratification framework \citep{frangakis2002principal, gilbert2008evaluating}, the proportion of treatment effect explained by the surrogate marker \citep{freedman1992statistical, wang2002measure, parast2016, parast2017evaluating, wang2020model, wang2021quantifying, wang2023robust}, and the relative and adjusted association to evaluate surrogacy in a setting where multiple clinical trials are available \citep{buyse1998criteria}. However, these methods can generally not handle the case when the surrogate is an intermediate \textit{outcome}, instead of an intermediately measured biomarker. In such a case, both the surrogate outcome and the primary outcome are subject to censoring. If we wish to make inference about the treatment effect at any particular intermediate point before the end of the trial, both the surrogate and primary outcome are somewhat ``missing" among individuals who have not yet experienced them. 
 
Certainly, some methods have been proposed to address this unique setting such as the survival methods of \citet{ghosh2008semiparametric, ghosh2009assessing} and \citet{lin1997estimating}, though they rely on complex joint modeling or other restrictive model assumptions. More recently, \citet{parast2020assessing} defined the proportion of treatment effect on the primary outcome that is explained by a censored surrogate outcome and estimated it semi-parametrically and non-parametrically. However, their assessment of surrogacy was limited to considering the restricted mean survival time (RMST) to quantify the treatment effect only, rather than the quantity that is typically of interest in clinical trials, the difference in overall survival, which limits the practical utility of their proposed approach. In addition, the metric proposed in  \citet{parast2020assessing} is not invariant to transformations of the surrogate and can change dramatically under certain transformations of the surrogate; in such a case, it would be ideal to instead consider how one may identify a transformation of the surrogate that is optimal in some sense. 

In this paper, we propose a non-parametric approach to evaluate a censored surrogate outcome, such as time to progression, when the primary outcome is also a censored time-to-event outcome, such as time to death, and the treatment effect of interest is the difference in overall survival. We define the proportion of the treatment effect on the primary outcome that is explained (PTE) by the censored surrogate outcome in this context, and estimate this proportion by defining and deriving an optimal transformation of the surrogate information. This framework can also be adapted to obtain the difference in RMST.

The remainder of the paper is organized as follows. In Section \ref{sec2}, we describe our setting and notation, define and derive an optimal transformation of the surrogate outcome information, and propose a PTE definition based on the transformation. In Section \ref{sims}, we conduct simulation studies to evaluate the finite sample performance of the proposed method. We then use our method to evaluate progression-free survival as a surrogate outcome for overall survival in an RCT among patients with metastatic colorectal cancer which compared chemotherapy plus Panitumumab vs. chemotherapy alone in Section \ref{app}. We provide concluding remarks in Section \ref{concluding} and include proofs of asymptotic results in the Appendix.

\section{Methods} \label{sec2}
\def\Dscr{\mathscr{D}}

\subsection{Setting and Notation}\label{sec21}

Let $T$ be the time to the primary outcome, death for example, $Y_t=I(T>t)$ be the primary outcome defined at a fixed time t, and $S$ be the time to an intermediate outcome such as time to myocardial infarction in a cardiovascular study or progression in a cancer study. The survival time $T$ is subject to censoring by the censoring time $C$. The intermediate outcome time $S$ is also subject to censoring  by $C$ but also subject to informative censoring by $T$. Under the standard causal inference framework, let $T\supa , C^{(a)}, S^{(a)} $  denote the potential survival time, potential censoring time, and potential intermediate outcome time under treatment $A=a, a=0, 1$. Note that the two sets of outcomes ($T\supone , C^{(1)}, S^{(1)} $) and ($T\supzero , C^{(0)}, S^{(0)} $) can not be observed simultaneously for one subject. We assume that we are in a single trial setting, that treatment is randomized, and without loss of generality, that treatment assignment is such that $P(A=1)=P(A=0)=1/2$. First, we focus on the treatment effect on the primary outcome quantified as the difference in survival at time $t$:
\begin{align}
\Delta(t)=E(Y_t^{(1)})-E(Y_t^{(0)})=\mu_1(t)-\mu_0(t),\ \text{where}\ Y_t^{(a)}=I(T^{(a)}>t),\ \mu_a(t)=E(Y_t^{(a)}). \label{treatmenteffect}
\end{align}
Our aim is to identify whether the surrogate can capture the treatment effect on the primary outcome. If there is no treatment effect on the outcome i.e., $\Delta(t)=0$, we argue that it would not be feasible or of interest to identify such a surrogate because the concept will then be ill-defined. Thus, we assume $\Delta(t)> 0$; note that the methods that follow are similarly applicable if $\Delta(t)< 0$, considering treatment as control and vice versa. 

\subsection{Surrogate Information and Proportion Explained}
We aim to evaluate to what degree the surrogate information collected up to some time $t_0\leq t$ can be used to make inference about the treatment effect on the primary outcome defined at $t$. Our approach will have four key features that, together, make this a unique contribution to the field; specifically, our approach will (1) be applicable to a setting where both the surrogate and primary outcome are time-to-event outcomes subject to censoring, (2) besides difference in overall survival, the framework can also be adapted to the difference in RMST as the treatment effect quantities of interest, and (3) be model-free in terms of definition and estimation.

To think about how we can utilize the surrogate information at $t_0$, it is helpful to consider two possible scenarios: (i) $T\leq t_0$ and (ii) $T>t_0$. Here, we are ignoring censoring in order to define the quantities of interest, but return to censoring when we do estimation in Section \ref{estimation}. In scenario (i), we already know the true primary outcome $Y_t=I(T>t)=0$; thus, we can simply focus on the true primary outcome in terms of the treatment effect regardless of the surrogate $S$. Under scenario (ii), there are two potential sub-cases: (iia) $S\leq t_0$, where the $S$ is observed to occur before $t_0$, and (iib)$S>t_0$, where $S$ does not occur before $t_0$ and thus, the value of $S$ is unknown at $t_0$. In summary, there are three situations: (i) $T\leq t_0$, (iia) $T>t_0\geq S$ and (iib) ($T>t_0, S>t_0$). Motivated by this breakdown, we define the surrogate information at $t_0$,  $Q_{t_0}$, as a combination of $\{I(T\leq t_0) 0,\ I(T >t_0) I( {S}\leq t_0)  {S},\ I(T>t_0) I( {S}> t_0)\}$. Next, we define a transformation, $g(\cdot)$, of $Q_{t_0}$ as
$$g(Q_{t_0})=I(T>t_0) \{I( {S}\leq t_0)  g_{1}( {S})+I( {S}>t_0) g_{2}\}.$$
With some abuse of notation, we sometimes drop $t$ and $t_0$ in $Y_t$ and $Q_{t_0}$ for clarity in notation below. 

Our goal is to find the optimal transformation function of the surrogate information at $t_0$, $g(S)=( g_{1}(S), g_{2})$, such that the treatment effect on this optimal transformation maximally explains the treatment effect on the primary outcome. This parallels the optimal transformation idea of \citet{wang2020model} but is further complicated by both censoring of the outcomes and the the fact that $Q_{t_0}$ is a collection of information rather than a single surrogate marker measurement. We derive the transformation function by minimizing
\begin{align}
L(g)=E\{Y^{(1)}-g(Q^{(1)})\}^2\ s.t. \ E\{Y^{(0)}-g(Q^{(0)})\}=0 \label{eq1} 
\end{align}
In Appendix \ref{appendixa}, we show that the solution to (\ref{eq1}) has the following forms:
\begin{align}
g_{1, opt}(s)&= \frac{\lambda  f_0(s, t_0, t_0)+ f_1(s, t, t_0)}{ f_1(s, t_0, t_0)},\\
g_{2, opt}&=\frac{\lambda P(T^{(0)}>t_0, S^{(0)}>t_0)+ P(T^{(1)}>t, S^{(1)}>t_0)}{P(T^{(1)}>t_0, S^{(1)}>t_0)},
\end{align}
where 
\begin{align*}
\lambda&=\bigg\{\int  \frac{f_0^2(s, t_0, t_0)}{f_1(s, t_0, t_0)}ds+\frac{P^2(T^{(0)}>t_0, S^{(0)}>t_0)}{P(T^{(1)}>t_0, S^{(1)}>t_0)}\bigg\}^{-1} \\
& \times  \bigg\{ \mu_0(t)-\int \frac{f_0(s,t_0, t_0)f_1(s, t, t_0)}{f_1(s, t_0, t_0)}ds-\frac{P(T^{(0)}>t_0, S^{(0)}>t_0)P(T^{(1)}>t, S^{(1)}>t_0)}{P(T^{(1)}>t_0, S^{(1)}>t_0)} \bigg\},
\end{align*}
$f_a(s, t, t_0)=P(T^{(a)}>t, S^{(a)} \leq t_0)  f_a(s| T>t, S\leq t_0)$ and $f_a(s| T>t, S\leq t_0)$ is the density of $S$ at $s$ given $(T>t, S\leq t_0, A=a)$. Thus, the transformation function we aim to find is 
$$\gopt(Q_{t_0})=I(T>t_0) \{I( {S}\leq t_0) g_{1, opt}( {S})+I( {S}>t_0) g_{2, opt}\}.$$

We define the PTE of this optimal transformation of the surrogate information as 
$$\PTE=\Delta_{\gopt(Q_{t_0})} /\Delta(t),$$
where $\Delta_{\gopt(Q_{t_0})}=E\{\gopt(Q_{t_0}^{(1)})-\gopt(Q_{t_0}^{(0)})\}$ is the treatment effect on $\gopt(Q_{t_0})$ and $\Delta(t)$ is the treatment effect on the primary outcome $Y_t$ defined in (\ref{treatmenteffect}).

\subsection{Estimation and Inference }\label{estimation}
The observed data for analysis consist of $n$ sets of independent and identically distributed random vectors $\Dscr = \{\bD_i = (X_i, \delta_i, A_i, I(X_i > t_0)I(S_i\leq t_0), I(X_i > t_0)I(S_i\leq t_0)S_i, I(X_i > t_0)I(S_i >t_0) ), i = 1, ..., n\}$, where  $T_i = T_i^{(1)}A_i + T_i^{(0)}(1-A_i)$, $C_i = C_i^{(1)}A_i + C_i^{(0)}(1-A_i)$,  $X_i=\min(T_i, C_i)$, $S_i = S_i^{(1)}A_i + S_i^{(0)}(1-A_i)$, and $C_i^{(a)}$ is assumed to be independent of $(T_i^{(a)}, S_i^{(a)})$ with $P(C_i^{(a)} > t)>0$ for $a=0,1$. We propose to estimate the unknown quantities nonparametrically as:
\begin{eqnarray*}
\muhat_a(t) & = &\frac{\sum_{i=1}^n \omegahat_{t,i}I(A_i =a)I(X_{i}>t)}{\sum_{i=1}^n \omegahat_{t,i}I(A_i=a)},\\
\hat{f}_a(s, t, t_0)&=& \frac{\sum_{i=1}^n I(A_i =a)K_h(S_i-s)I(X_i>t, S_i \leq t_0)\omegahat_{t,i}}{\sum_{i=1}^n I(A_i=a) \omegahat_{t,i}}, \\
\hat{P}(T^{(a)}>t, S^{(a)}>t_0)&=&\frac{\sum_{i=1}^n I(A_i =a) I(X_i>t, S_i>t_0)\omegahat_{t,i}}{\sum_{i=1}^n I(A_i =a) \omegahat_{t,i}},
\end{eqnarray*}
where $\omegahat_{t,i}=\{I(X_i\leq t)\delta_i+I(X_i > t)\}/\Ghat_{A_i}(X_i \wedge t)$ is the weight accounting for censoring, $\Ghat_{a}(\cdot)$ is the Kaplan-Meier estimator of $G_{a}(\cdot) = P(C > t \mid A = a)$, and $K_{h}(\cdot)=K(\cdot/h)/h$, $K(\cdot)$ is a symmetric kernel function with bandwidth $h$. Correspondingly, we get the estimators
\begin{align*}
\ghat_{1}(s)&=\frac{\hat{\lambda}  \hat{f}_0(s, t_0, t_0)+ \hat{f}_1(s, t, t_0)}{ \hat{f}_1(s, t_0, t_0)},\\
\ghat_{2}&=\frac{\hat{\lambda }\hat{P}(T^{(0)}>t_0, S^{(0)}>t_0)+ \hat{P}(T^{(1)}>t, S^{(1)}>t_0)}{\hat{P}(T^{(1)}>t_0, S^{(1)}>t_0)},\\
\text{where}\ \hat{\lambda}&=\bigg\{\int  \frac{\hat{f}_0^2(s, t_0, t_0)}{\hat{f}_1(s, t_0, t_0)}ds+\frac{\hat{P}^2(T^{(0)}>t_0, S^{(0)}>t_0)}{\hat{P}(T^{(1)}>t_0, S^{(1)}>t_0)}\bigg\}^{-1} \\
& \times  \bigg\{ \muhat_0(t)-\int \frac{\hat{f}_0(s,t_0, t_0)\hat{f}_1(s, t, t_0)}{\hat{f}_1(s, t_0, t_0)}ds-\frac{\hat{P}(T^{(0)}>t_0, S^{(0)}>t_0)\hat{P}(T^{(1)}>t, S^{(1)}>t_0)}{\hat{P}(T^{(1)}>t_0, S^{(1)}>t_0)} \bigg\}.
\end{align*}
The estimator of $\gopt(Q_{t_0,i})$ is 
$$\ghat_i:=I(X_i>t_0) \{I( {S_i}\leq t_0) \ghat_{1}( {S_i})+I( {S_i}>t_0) \ghat_{2}\},$$
and thus, our estimate of the PTE is 
$$\PTEhat=\Deltahat_{\ghat}/\Deltahat(t),$$
where 
$$\Deltahat(t)  = \muhat_1(t) - \muhat_0(t),  \qquad \quad \Deltahat_{\ghat } = \muhat_{\ghat,1} - \muhat_{\ghat,0},$$
$$ \muhat_a(t)  = \frac{\sum_{i=1}^n \omegahat_{t,i}I(A_i =a)I(X_{i}>t)}{\sum_{i=1}^n \omegahat_{t,i}I(A_i=a)}, \qquad
\muhat_{\ghat,a} =\frac{\sum_{i=1}^n \omegahat_{t_0,i}I(A_i=a) \ghat_i }{\sum_{i=1}^n \omegahat_{t_0,i}I(A_i=a) }.$$


In Appendix \ref{appendixb} of the Supplementary Materials, we show that under the conditions (C1)-(C4) in Appendix \ref{appendixb}, $\PTE$ is between 0 and 1. Using similar strategies to that of \citet{wang2020model}, it can be shown that $\PTEhat$ is a consistent estimator of $\PTE$, and when $h = O(n^{-\nu})$ with $\nu \in (1/4,1/2)$, $\nhalf(\PTEhat - \PTE)$ is asymptotically normal with a complicated form of the asymptotic variance. In practice, we estimate the asymptotic variances via resampling similar to those employed in \citet{parast2016}. In simulation studies following, we chose $K(\cdot)$ as a Gaussian kernel with bandwidth $h=h_{opt}n^{-c_0}, c_0=0.06$, where $h_{opt}$ is found in \citet{scott1992multivariate}.

\subsection{Surrogate Value}\label{sec24}
Notably, our definition of $Q_{t_0}$ involves primary outcome information. We argue that this is absolutely reasonable because, after all, the primary outcome is the primary outcome and if it actually occurs before $t_0$ then we, of course, already know that $T\leq t$. However, it is reasonable to ask whether the $\PTE$ is actually impacted by the surrogate itself, or the primary outcome. That is, what is the value of the surrogate information alone in terms of the $\PTE$? To answer this question, we compare the above proposed method to a method that does not use the information from the surrogate $S$ at $t_0$. That is, to a method that only uses the information from the primary outcome, $T$, alone at $t_0$. For the purpose of this definition, with a slight abuse of notation, here we define $Q_{t_0}^*$ at $t_0$ as a combination of $\{I(T\leq t_0) 0,\ I(T >t_0) I( {T}\leq t_0)  T,\ I(T>t_0) I( {T}> t_0)\}$, and a transformation of $Q_{t_0}^*$ as $g(Q_{t_0}^*)=I(T>t_0) \{I( {T}\leq t_0)  g^*_{1}( {T})+I( T>t_0) g^*_{2}\}=(T>t_0)g^*_{2}.$ Notice that there is no $S$ in the preceding sentence. Similar to the derivation of (\ref{eq1}), the optimal transformation of the surrogate information at $t_0$ is
\begin{align*}
g^*_{2, opt}&=\frac{\lambda P(T^{(0)}>t_0)+ P(T^{(1)}>t)}{P(T^{(1)}>t_0)},\\
\text{where}\ 
\lambda&=\bigg\{\frac{P^2(T^{(0)}>t_0)}{P(T^{(1)}>t_0)}\bigg\}^{-1}  \bigg\{ \mu_0(t)-\frac{P(T^{(0)}>t_0)P(T^{(1)}>t)}{P(T^{(1)}>t_0)} \bigg\}.
\end{align*}
Thus, the optimal transformation function and the corresponding $\PTE$, denoted as $\PTE_{Ind}$, are
\begin{align*}
\gopt(Q_{t_0}^*)&=I(T>t_0)  g^*_{2, opt},\\
\PTE_{Ind}&=\Delta_{\gopt(Q_{t_0}^*)} /\Delta(t).
\end{align*}
They can be estimated similarly to Section \ref{estimation}. The difference $\PTE-\PTE_{Ind}$ indicates the added value of the surrogate itself.

\def\subI{_{\mbox{\tiny I}}}
\def\subII{_{\mbox{\tiny II}}}
\def\subIII{_{\mbox{\tiny III}}}
\def\Gsc{\mathcal{G}}

\section{Simulation Study \label{sims}}
Simulation studies were conducted to evaluate the finite sample performance of the proposed method and compare with existing method of \citet{parast2020assessing}. But since the outcome of interest in \citet{parast2020assessing} is RMST, we further define a $\PTE^{rmst}$ based on the proposed optimal transformation as follows so that $\PTE^{rmst}$ is comparable to the PTE of \citet{parast2020assessing}.

The restricted survival time by time $\tau$ is $\min\{T,\tau\}=\int_0^{\tau} I(T>t) dt$ and the corresponding quantity with the optimal transformation is $G_{\tau}(t_0)=\int_0^{t_0} I(T>t) dt+\int_{t_0}^{\tau} \gopt(Q_{t_0},t) dt$, where the transformation is actually a function of $(S, t_0, t)$, or a function of $(Q_{t_0},t)$, $\gopt(Q_{t_0},t)=I(T>t_0) \{I( {S}\leq t_0) g_{1, opt}(S,t)+I( {S}>t_0) g_{2, opt}(t)\}.$ The treatment effect on the restricted survival time is 
$$\Delta^{rst}_{\tau}=E\left[\int_0^{\tau} I(T^{(1)}>t) dt-\int_0^{\tau} I(T^{(0)}>t) dt\right]=\int_0^{\tau} \Delta(t) dt,$$ and treatment effect on $G_{\tau}(t_0)$ is
\begin{align*}
\Delta_{G_{\tau}(t_0)}&=E\left[\int_0^{t_0} I(T^{(1)}>t) dt-\int_{t_0}^{\tau} I(T^{(0)}>t) dt\right]+E\left[\int_{t_0}^{\tau}\gopt(Q_{t_0}^{(1)},t)dt-\int_{t_0}^{\tau}\gopt(Q_{t_0}^{(0)},t)dt\right]\\
&=\int_0^{t_0} \Delta(t) dt+\int_{t_0}^{\tau} \Delta_{\gopt(Q_{t_0},t)} dt.
\end{align*}
The proportion of the treatment effect, as quantified by the difference in RMST, that is explained by the surrogate information can be defined correspondingly as 
$$\PTE^{rsmt}(t_0,t)=\Delta_{G_{t}(t_0)} /\Delta^{rst}_{t}.$$
Similarly we can define $\PTE^{rsmt}_{Ind}(t_0,t)$ using the transformation function in Section \ref{sec24} and evaluate the added value of the surrogate S at $t_0$ based on $\PTE^{rsmt}(t_0,t)-\PTE^{rsmt}_{Ind}(t_0,t)$. These quantities can be estimated similarly to Section \ref{estimation}. 

For all settings, the sample size was $n=2000$, with 1000 for each treatment group, variances were estimated using the perturbation resampling method \citep{parast2016} based on 500 replications and all results were summarized based on 500 simulated datasets. In addition, for all settings, we let $t = 5$, and generated censoring as $C^{(a)} \sim \text{exponential}(0.12)$ in both groups. Note that information on $S^{(a)}$ is available only for those with $X^{(a)} > S^{(a)}$. We examined results when $t_0$ was 1, 2, or 3, with the expectation that when $t_0$ is closer to $t$, the PTE will be closer to 1. 

In the first setting, setting (1), we generated $S^{(1)}, S^{(0)}, T^{(1)}, T^{(0)}$ as:
\begin{alignat*}{2}
S\supone&\sim  Weibull(shape = 1, scale = 6),\\
S\supzero&\sim  Weibull(shape = 1, scale = 4),\\
T\supzero&=-\log(1-U^{(1)})\  5S^{(1)}, \text{where}\ U^{(1)} \sim Uniform(0,1),\\
T\supone&= -\log(1-U^{(0)})\ 3S^{(0)}, \text{where}\ U^{(0)} \sim Uniform(0,1).
\end{alignat*}
The overall censoring rate was approximately 58\%, 66\% for group 1 ($A=1$) and 49\% for group 0 (A=0), respectively.  In the second setting, setting (2), we generated $S^{(1)}, S^{(0)}, T^{(1)}, T^{(0)}$ as:
\begin{alignat*}{2}
S\supone&\sim  Exp(0.6),\\
S\supzero&\sim  Exp(2),\\
T\supzero&=S^{(1)} + E^{(1)} + \exp{(N^{(1)})}, \text{where}\ E^{(1)} \sim Exp(1/8),\ N^{(1)} \sim N(0, 0.1^2),\\
T\supone&=S^{(0)} + E^{(0)} + \exp{(N^{(0)})}, \text{where}\ E^{(0)} \sim Exp(1/4),\ N^{(0)} \sim N(0, 0.1^2).
\end{alignat*}
The censoring rate was approximately 53\%, 63\% for group 1 ($A=1$) and 43\% for group 0 (A=0), respectively. These two settings were chosen so that the added value of the surrogate to the PTE was relatively minor in setting (1) but relative large in setting (2), following \citet{parast2020assessing}. In the third setting, setting (3), we generated $S^{(1)}, S^{(0)}, T^{(1)}, T^{(0)}$ as:
\begin{alignat*}{2}
S\supone&\sim  Exp(0.6),\\
S\supzero&\sim  Exp(2),\\
T\supzero&=S^{(1)} -\log S^{(1)}+ E^{(1)} + \exp{(N^{(1)})}, \text{where}\ E^{(1)} \sim Exp(1/4),\ N^{(1)} \sim N(0, 0.1^2),\\
T\supone&=S^{(0)}-\log S^{(0)} + E^{(0)} + \exp{(N^{(0)})}, \text{where}\ E^{(0)} \sim Exp(1/2),\ N^{(0)} \sim N(0, 0.1^2).
\end{alignat*}
The censoring rate was approximately 47\%, 51\% for group 1 ($A=1$) and 42\% for group 0 (A=0), respectively. In this setting, the outcome and the surrogate are not monotonically related, an assumption required by \citet{parast2020assessing}. The purpose of this setting was to assess how our proposed approach handled a violation of this assumption.

Simulation results are summarized in Table \ref{tab1}, \ref{tab2} and \ref{tab3}. The proposed estimates (for either PTE or $g_{2}$) has negligible bias and the average of standard error estimates (ASE) is close to the corresponding empirical standard error (ESE). The empirical coverage probability (CP) is close to the nominal level 95\%. Generally, as $t_0$ increases, the $\PTE$ estimate increases, indicating a higher surrogacy of later year surrogate information for 5-year survival, as expected.
Tables \ref{tab1}, \ref{tab2} and \ref{tab3} also show the estimate of $\PTE_{Ind}$ described in Section \ref{sec24}. Results show that the estimates of $\PTE$ are generally higher than the corresponding $\PTE_{Ind}$ estimates, reflecting the added value of the actual surrogate $S$ at $t_0$. The added value of the surrogate in setting (1) is small while those in settings (2) and (3) are relatively large. 

From Tables \ref{tab1}, \ref{tab2} and \ref{tab3}  we can also see that the $\PTE^{rmst}$ estimates are generally close to the $\PTE$ estimates of \citet{parast2020assessing}, denoted as $\PTE_{R}$. However, in setting (3), both estimates are negative when $t_0=1$ and 2, so are hard to explain. This may be because the assumptions for the validity of these estimators are not satisfied. 
 
 \def\CDfn{\mbox{CD4}_0}

\section{Application to the Panitumumab Randomized Trial\label{app}}

\begin{figure}[!h]
\centering
\includegraphics[width=.7\textwidth]{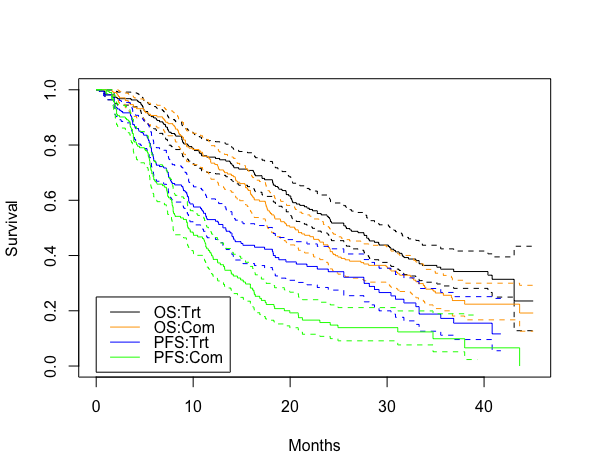}
\caption{Kaplan–Meier Estimates of Overall Survival (OS) and Progression-free Survival (PFS) with 95\% confidence intervals, where Trt denotes Panitumumab+FOLFOX4 and Com denotes  FOLFOX4.}\label{fig1}
\end{figure}

We used our proposed approach to evaluate progression-free survival as a surrogate outcome for overall survival in an RCT among patients with metastatic colorectal cancer which compared chemotherapy plus Panitumumab vs. chemotherapy alone. Specifically, the Panitumumab Randomized Trial in Combination with Chemotherapy for Metastatic Colorectal Cancer to Determine Efficacy (PRIME) compared the efficacy and safety of panitumumab–FOLFOX4 with those of FOLFOX4 alone in the first-line treatment of patients. The study began on August 1, 2006 with follow up to August 1, 2009, where 54$\%$ of the patients had died. In our illustration, we specifically focus on the 424 participants who were identified at baseline as having tumors with non-mutated RAS (no KRAS or NRAS mutations in exons 2, 3, or 4). Among these participants, it has been shown that panitumumab–FOLFOX4, as compared with FOLFOX4 alone, was associated with a significant improvement in progression-free survival and a significant improvement in overall survival \citep{douillard2013panitumumab}, as seen from the Kaplan-Meier curves in Figure \ref{fig1}. Our goal was to investigate to what extent the surrogate information at $t_0=6, 10, 14, ..., 34$ months captures the treatment effect on overall survival at $t=36$ months, using the proposed method.  

The proposed estimates of the proportion of treatment effect explained by the surrogate are shown in Table \ref{tab4}. Results show that for $t_0$ greater than 14 months, the surrogate information is capturing more than 50\% of the overall treatment effect. Prior work has suggested considering a surrogate marker or outcome a “good” surrogate if the lower bound of the 95$\%$ confidence interval is above some threshold such as 0.50, rather than simply the point estimate \citep{lin1997estimating}. Thus, we also show this lower bound at each $t_0$. At $t_0 = 26$ months, for example, the estimated $\PTE$ is 0.98 and the lower bound is $0.62$. Table \ref{tab4} includes estimates of $\PTE_{Ind}$ which, compared to the proposed $\PTE$ estimates, tend to be quite a bit lower implying additional value of the surrogate outcome information i.e., progression. Estimates of $\PTE^{rmst}$, $\PTE^{rmst}_{Ind}$, and the estimate of \citet{parast2020assessing}, $\PTE_{R}$ show that $\PTE^{rmst}$ estimates were generally higher than the proposed $\PTE$ estimates, which were generally higher than $\PTE_{R}$ estimates; results show some negative values for early time points, likely due to a brief period of time where the treated survival curve was below the control group survival curve, shown in Figure \ref{fig1}. In general, all the estimators show a similar surrogacy trend as $t_0$ increases to $t$ with evidence that for some time points, progression-free survival captures a substantial amount of the treatment effect on overall survival. 

\section{Discussion \label{concluding}}

In this paper, we proposed a novel statistical method to evaluate a censored surrogate outcome when the primary outcome is also a censored time-to-event outcome by defining and deriving an optimal transformation of the surrogate information at an earlier time point, $t_0$, and the proportion of the treatment effect explained by this optimal transformation. The three key features of our approach i.e., that it is applicable to a setting where both the surrogate and primary outcome are time-to-event outcomes subject to censoring, that the $\PTE$ of interest is the difference in overall survival and the optimal transformation can also be used to derive the difference in RMST as the treatment effect, and are model-free in terms of definition and estimation, highlight the utility of this method in practice.  Our numerical studies demonstrated good performance of the proposed method, and our application to the PRIME trial showed the surrogate value of progression-free survival as a surrogate outcome for overall survival.

There exist further extensions of this approach that could be considered for future methodological development. For example, in our application in the PRIME trial, the surrogate could be alternatively considered as longitudinal surrogate information and one may consider evaluating it as such, instead of at a single point in time, $t_0$. In addition, we defined the surrogate information $Q_{t_0}$ a combination of $\{I(T\leq t_0) 0,\ I(T >t_0) I( {S}\leq t_0)  {S},\ I(T>t_0) I( {S}> t_0)\}$ and thus, the observation of S was only utilized when $T> t_0$ and $S\leq t_0$. Instead, one may consider using the observation of S even when $T \leq  t_0$, which may possibly be useful.

For subgroups of patients with different characteristics, i.e., ancestries, the utility of the surrogate may be different. Thus, surrogate evaluation results using the whole population may be dominated by the group of individuals with the largest proportion, e.g., European-origin individuals. To address inequalities in the representation of different population in the whole population, it will be important for future work to consider subgroup-specific PTE measure, or a covariate-specific measure of PTE.

%

\bibliographystyle{apalike}
\bibliography{ref}

\begin{thebibliography}{}

\bibitem[Alberti et~al., 2006]{alberti2006metabolic}
Alberti, K. G. M.~M., Zimmet, P., and Shaw, J. (2006).
\newblock Metabolic syndrome—a new world-wide definition. a consensus
  statement from the international diabetes federation.
\newblock {\em Diabetic medicine}, 23(5):469--480.

\bibitem[Buyse and Molenberghs, 1998]{buyse1998criteria}
Buyse, M. and Molenberghs, G. (1998).
\newblock Criteria for the validation of surrogate endpoints in randomized
  experiments.
\newblock {\em Biometrics}, 54(3):1014--1029.

\bibitem[Douillard et~al., 2013]{douillard2013panitumumab}
Douillard, J.-Y., Oliner, K.~S., Siena, S., Tabernero, J., Burkes, R., Barugel,
  M., Humblet, Y., Bodoky, G., Cunningham, D., Jassem, J., et~al. (2013).
\newblock Panitumumab--folfox4 treatment and ras mutations in colorectal
  cancer.
\newblock {\em New England Journal of Medicine}, 369(11):1023--1034.

\bibitem[FDA, 2022]{FDA_approved}
FDA (2022).
\newblock Table of surrogate endpoints that were the basis of drug approval or
  licensure.
\newblock
  {https://www.fda.gov/drugs/development-resources/table-surrogate-endpoints-were-basis-drug-approval-or-licensure}.

\bibitem[Food et~al., 2008]{us2008guidance}
Food, U., Administration, D., et~al. (2008).
\newblock Guidance for industry: diabetes mellitus: developing drugs and
  therapeutic biologics for treatment and prevention.
\newblock {\em Services UDoHaH, ed}.

\bibitem[Frangakis and Rubin, 2002]{frangakis2002principal}
Frangakis, C.~E. and Rubin, D.~B. (2002).
\newblock Principal stratification in causal inference.
\newblock {\em Biometrics}, 58(1):21--29.

\bibitem[Freedman et~al., 1992]{freedman1992statistical}
Freedman, L.~S., Graubard, B.~I., and Schatzkin, A. (1992).
\newblock Statistical validation of intermediate endpoints for chronic
  diseases.
\newblock {\em Statistics in {M}edicine}, 11(2):167--178.

\bibitem[Ghosh, 2008]{ghosh2008semiparametric}
Ghosh, D. (2008).
\newblock Semiparametric inference for surrogate endpoints with bivariate
  censored data.
\newblock {\em Biometrics}, 64(1):149--156.

\bibitem[Ghosh, 2009]{ghosh2009assessing}
Ghosh, D. (2009).
\newblock On assessing surrogacy in a single trial setting using a
  semicompeting risks paradigm.
\newblock {\em Biometrics}, 65(2):521--529.

\bibitem[Gilbert and Hudgens, 2008]{gilbert2008evaluating}
Gilbert, P.~B. and Hudgens, M.~G. (2008).
\newblock Evaluating candidate principal surrogate endpoints.
\newblock {\em Biometrics}, 64(4):1146--1154.

\bibitem[Group et~al., 2000]{allhat2000major}
Group, A. C.~R. et~al. (2000).
\newblock Major cardiovascular events in hypertensive patients randomized to
  doxazosin vs chlorthalidone: the antihypertensive and lipid-lowering
  treatment to prevent heart attack trial (allhat).
\newblock {\em Jama}, 283:1967--1975.

\bibitem[Lin et~al., 1997]{lin1997estimating}
Lin, D., Fleming, T., De~Gruttola, V., et~al. (1997).
\newblock Estimating the proportion of treatment effect explained by a
  surrogate marker.
\newblock {\em Statistics in medicine}, 16(13):1515--1527.

\bibitem[Lorenzo et~al., 2003]{lorenzo2003metabolic}
Lorenzo, C., Okoloise, M., Williams, K., Stern, M.~P., and Haffner, S.~M.
  (2003).
\newblock The metabolic syndrome as predictor of type 2 diabetes: the san
  antonio heart study.
\newblock {\em Diabetes care}, 26(11):3153--3159.

\bibitem[Parast et~al., 2017]{parast2017evaluating}
Parast, L., Cai, T., and Tian, L. (2017).
\newblock Evaluating surrogate marker information using censored data.
\newblock {\em Statistics in medicine}, 36(11):1767--1782.

\bibitem[Parast et~al., 2016]{parast2016}
Parast, L., McDermott, M.~M., and Tian, L. (2016).
\newblock Robust estimation of the proportion of treatment effect explained by
  surrogate marker information.
\newblock {\em Statistics in {M}edicine}, 35(10):1637--1653.

\bibitem[Parast et~al., 2020]{parast2020assessing}
Parast, L., Tian, L., and Cai, T. (2020).
\newblock Assessing the value of a censored surrogate outcome.
\newblock {\em Lifetime data analysis}, 26:245--265.

\bibitem[Ridker et~al., 2005]{ridker2005randomized}
Ridker, P.~M., Cook, N.~R., Lee, I.-M., Gordon, D., Gaziano, J.~M., Manson,
  J.~E., Hennekens, C.~H., and Buring, J.~E. (2005).
\newblock A randomized trial of low-dose aspirin in the primary prevention of
  cardiovascular disease in women.
\newblock {\em New England Journal of Medicine}, 352(13):1293--1304.

\bibitem[Robins and Greenland, 1992]{robins1992identifiability}
Robins, J.~M. and Greenland, S. (1992).
\newblock Identifiability and exchangeability for direct and indirect effects.
\newblock {\em Epidemiology}, pages 143--155.

\bibitem[Scott, 1992]{scott1992multivariate}
Scott, D. (1992).
\newblock {\em Multivariate density estimation}.
\newblock John Wiley \& Sons.

\bibitem[Wang et~al., 2021]{wang2021quantifying}
Wang, X., Cai, T., Tian, L., Bourgeois, F., and Parast, L. (2021).
\newblock Quantifying the feasibility of shortening clinical trial duration
  using surrogate markers.
\newblock {\em Statistics in medicine}, 40(28):6321--6343.

\bibitem[Wang et~al., 2023]{wang2023robust}
Wang, X., Parast, L., Han, L., Tian, L., and Cai, T. (2023).
\newblock Robust approach to combining multiple markers to improve surrogacy.
\newblock {\em Biometrics}, 79(2):788--798.

\bibitem[Wang et~al., 2020]{wang2020model}
Wang, X., Parast, L., Tian, L., and Cai, T. (2020).
\newblock Model-free approach to quantifying the proportion of treatment effect
  explained by a surrogate marker.
\newblock {\em Biometrika}, 107(1):107--122.

\bibitem[Wang and Taylor, 2002]{wang2002measure}
Wang, Y. and Taylor, J.~M. (2002).
\newblock A measure of the proportion of treatment effect explained by a
  surrogate marker.
\newblock {\em Biometrics}, 58(4):803--812.

\end{thebibliography}

\newpage

\begin{table}[ht]
\centering
\begin{tabular}{r|rrrr|rr}
  \hline
$t_0$  & $\PTE_{true}$ & Est & ESE$_{\mbox{\tiny ASE}}$ & CP& $\PTE_{Ind}$ & ESE    \\ 
  \hline
1 &.350 & .363 & $.062_{.075}$ & .986 & .311 & .056  \\ 
   \hline
2  &.594 & .582 & $.077_{.086}$ & .958 & .511 & .073\\ 
   \hline
3 &.759 & .751 & $.082_{.081}$ & .922 & .694 & .080 \\ 
   \hline
\end{tabular}

\ \

\centering
\begin{tabular}{r|rr|rr|rr}
  \hline
$t_0$  & $\PTE^{rmst}$ &   ESE   &$\PTE_{Ind}^{rmst}$ & ESE& $\PTE_{R}$ & ESE    \\ 
  \hline
1  & .586 & .057 & .544 & .060 & .586 & .070 \\ 
   \hline
2  &.788 & .055 & .745 & .055& .809 & .057\\ 
   \hline
3 &.905 & .040 & .877 & .041 & .926 & .047 \\ 
   \hline
\end{tabular}

\ \

\centering
\begin{tabular}{rrrrrr}
  \hline
$t_0$ & $g_{2, true}$ & Est & ESE$_{\mbox{\tiny ASE}}$ & CP \\ 
  \hline
1 & .684 & .689 & $.020_{.021}$ & .946 \\ 
   \hline
2 &.806 & .809 & $.025_{.024}$ & .935 \\ 
   \hline
3 &.897 & .892 & $.024_{.025}$ & .954 \\ 
   \hline
\end{tabular}
\caption{Estimates (Est) of $g_{2}$, $\PTE$ and $\PTE_{Ind}$, on overall survival, $\PTE^{rmst}$, $\PTE^{rmst}_{Ind}$ and $\PTE_{R}$, on restricted survival time, along with their empirical standard errors (ESE) under settings (1) with $n=2000$; for our proposed PTE estimates,
we also present the average of the estimated standard errors (ASE, shown in subscript) along with the empirical coverage probabilities (CP) of the 95\% confidence intervals. } \label{tab1}
\end{table}

 \newpage
 

\begin{table}[ht]
\centering
\begin{tabular}{r|rrrr|rr}
  \hline
$t_0$  & $\PTE_{true}$ & Est & ESE$_{\mbox{\tiny ASE}}$ & CP& $\PTE_{Ind}$ & ESE    \\ 
  \hline
1 &.554 & .534 & $.055_{.056}$ & .929 & .001 & .002  \\ 
   \hline
2  &.608 & .589 & $.050_{.053}$ & .942 & .150 & .023\\ 
   \hline
3 &.713 & .692 & $.045_{.051}$ & .958 & .407 & .047  \\ 
   \hline
\end{tabular}

\ \

\centering
\begin{tabular}{r|rr|rr|rr}
  \hline
$t_0$  & $\PTE^{rmst}$ &   ESE   &$\PTE_{Ind}^{rmst}$ & ESE& $\PTE_{R}$ & ESE    \\ 
  \hline
1  & .342 & .051 & .004 & .006& .290 & .059\\ 
   \hline
2  &.617 & .042 & .383 & .040 & .553 & .055\\ 
   \hline
3 &.811 & .030 & .716 & .037 & .819 & .036 \\ 
   \hline
\end{tabular}

\ \

\centering
\begin{tabular}{rrrrrr}
  \hline
$t_0$ & $g_{2, true}$ & Est & ESE$_{\mbox{\tiny ASE}}$ & CP \\ 
  \hline
1 &.792 & .799 & $.031_{.030}$ & .926 \\ 
   \hline
2 & .901 & .898 & $.038_{.039}$ & .944 \\ 
   \hline
3 &.977 & .969 & $.045_{.046}$ & .950 \\ 
   \hline
\end{tabular}
\caption{Estimates (Est) of $g_{2}$, $\PTE$ and $\PTE_{Ind}$, on overall survival, $\PTE^{rmst}$, $\PTE^{rmst}_{Ind}$ and $\PTE_{R}$, on restricted survival time, along with their empirical standard errors (ESE) under settings (2) with $n=2000$; for our proposed PTE estimates,
we also present the average of the estimated standard errors (ASE, shown in subscript) along with the empirical coverage probabilities (CP) of the 95\% confidence intervals. } \label{tab2}
\end{table}

\newpage
\begin{table}[ht]
\centering
\begin{tabular}{r|rrrr|rr}
  \hline
$t_0$  & $\PTE_{true}$ & Est & ESE$_{\mbox{\tiny ASE}}$ & CP& $\PTE_{Ind}$ & ESE    \\ 
  \hline
1 &.356 & .318 & $.080_{.094}$ & .976 & .000 & .000  \\ 
   \hline
2  &.373 & .341 & $.078_{.093}$ & .969 & .002 & .004\\ 
   \hline
3 &.490 & .436 & $.079_{.092}$ & .952 & .169 & .042  \\ 
   \hline
\end{tabular}

\ \

\centering
\begin{tabular}{r|rr|rr|rr}
  \hline
$t_0$  & $\PTE^{rmst}$ &   ESE   &$\PTE_{Ind}^{rmst}$ & ESE& $\PTE_{R}$ & ESE    \\ 
  \hline
1  &-.367 & .206 & .000 & .000 & -.174 & .103 \\ 
   \hline
2  &-.067 & .134 & .010 & .021 & -.085 & .085\\ 
   \hline
3 &.429 & .103 & .481 & .072 & .434 & .096\\ 
   \hline
\end{tabular}

\ \

\centering
\begin{tabular}{rrrrrr}
  \hline
$t_0$ & $g_{2, true}$ & Est & ESE$_{\mbox{\tiny ASE}}$ & CP \\ 
  \hline
1 &.575 & .568 & $.031_{.031}$ & .951  \\ 
   \hline
2 & .667 & .666 & $.045_{.043}$ & .944 \\ 
   \hline
3 &.778 & .775 & $.052_{.055}$ & .950 \\ 
   \hline
\end{tabular}
\caption{Estimates (Est) of $g_{2}$, $\PTE$ and $\PTE_{Ind}$, on overall survival, $\PTE^{rmst}$, $\PTE^{rmst}_{Ind}$ and $\PTE_{R}$, on restricted survival time, along with their empirical standard errors (ESE) under settings (3) with $n=2000$; for our proposed PTE estimates,
we also present the average of the estimated standard errors (ASE, shown in subscript) along with the empirical coverage probabilities (CP) of the 95\% confidence intervals. } \label{tab3}
\end{table}

\newpage

\ \

\begin{table}[ht]
\centering
\begin{tabular}{rrrrrrrrrrrrrrr}
  \hline
$t_0$ & $\PTE$ & $\PTE_{Ind}$  & $\PTE^{rmst}$& $\PTE^{rmst}_{Ind}$ & $\PTE_{R}$& $\Low$ & $\Low^{rmst}$ & $\Low_R$ \\ 
  \hline
    6 & 0.20 (0.20) & -0.03 (0.12) & 0.29 (0.23) & -0.08 (0.44) & 0.15 (0.45) & -0.19 & -0.15 & -0.74 \\ 
    10 & 0.30 (0.21) & 0.00 (0.24) & 0.23 (0.22) & 0.01 (0.35) & -0.13 (0.57) & -0.10 & -0.19 & -1.25 \\ 
    14 & 0.56 (0.22) & 0.13 (0.18) & 0.48 (0.23) & 0.26 (0.31) & 0.03 (0.54) & 0.13 & 0.04 & -1.02 \\ 
    18 & 0.84 (0.18) & 0.55 (0.17) & 1.01 (0.13) & 0.88 (0.17) & 0.67 (0.31) & 0.48 & 0.76 & 0.06 \\ 
    22 & 0.77 (0.21) & 0.51 (0.25) & 0.90 (0.15) & 0.77 (0.30) & 0.71 (0.25) & 0.36 & 0.60 & 0.23 \\ 
    26 & 0.98 (0.18) & 0.79 (0.22) & 1.19 (0.19) & 1.14 (0.76) & 0.86 (0.17) & 0.62 & 0.82 & 0.53 \\
    30 & 0.79 (0.20) & 0.59 (0.28) & 0.99 (0.20) & 0.86 (0.29) & 0.76 (0.19) & 0.39 & 0.60 & 0.39 \\
    34 & 1.02 (0.17) & 0.87 (0.26) & 1.00 (0.13) & 0.97 (0.16) & 0.86 (0.12) & 0.70 & 0.74 & 0.61 \\ 
   \hline
\end{tabular}
\caption{Estimates of $\PTE$, $\PTE_{Ind}$, $\PTE^{rmst}$, $\PTE^{rmst}_{Ind}$ and $\PTE_{R}$; the numbers in the brackets are the estimated standard errors; $\Low$, $\Low^{rmst}$, and $\Low_{R}$ are the lower bounds of the 95$\%$ confidence intervals for $\PTE$, $\PTE^{rmst}$, and $\PTE_R$, respectively.} \label{tab4}
\end{table}

\newpage 
\appendix

\begin{appendices}
\def\gtildeopt{\widetilde{g}_{\mbox{\tiny opt}}}

\section{} \label{appendixa}

In this section, we derive the specific form for the optimal transformation function of the surrogate information, $\gopt(\cdot)=(g_{1, opt}(\cdot),g_{2, opt})$.
We aim to solve the following problem:
\begin{align*}
\min_g L(g)=E\{Y^{(1)}-g(Q^{(1)})\}^2, \quad \mbox{given}\ E\{Y^{(0)}-g(Q^{(0)})\}=0.
\end{align*}
Since
\begin{align*}
Y-g(Q)&=I(T>t)-I(T>t_0)\ \{I( {S}\leq t_0)  g_{1}( {S})+I( {S}>t_0) g_{2}\}\\
&=I(T>t_0)\{I(T>t)-I( {S}\leq t_0)  g_{1}( {S})-I( {S}>t_0) g_{2}\},
\end{align*}
we have
\begin{align*}
L(g)&=E\{Y^{(1)}-g(Q^{(1)})\}^2\\
&=E \left[ I(T^{(1)}>t)+ I(T^{(1)}>t_0)I( {S}^{(1)}\leq t_0) g^2_2( {S}^{(1)})+I(T^{(1)}>t_0)I( {S}^{(1)}>t_0) g^2_{3}\right]\\
&-2E\left[I(T^{(1)}>t)\{I( {S}^{(1)}\leq t_0)  g_{1}( {S}^{(1)})+I( {S}^{(1)}>t_0) g_{2}\}\right]\\
&=\mu_1(t)\!+P(T^{(1)}>t_0, S^{(1)} \leq t_0)\!\int \!g^2_2(s)f_1(s| T>t_0, S \leq t_0)ds+\!P(T^{(1)}\!>\!t_0, S^{(1)}\!>\!t_0) g^2_3\\
&-\!2 P(T^{(1)}>t, S^{(1)} \leq t_0) \int  g_{1}(s)f_1(s| T>t, S \leq t_0)ds\!-\!2P(T^{(1)}\!>\!t, S^{(1)}\!>\!t_0) g_{2}.
\end{align*}
Our optimization problem is thus,
$$\min_{g} \Lsc( g), \quad \mbox{given that} \quad \Gbb( g) = \mu_0(t),$$
where we used the functional notation
\begin{align*}
\Lsc( g) &= P(T^{(1)}>t_0, S^{(1)} \leq t_0)\!\int \!g^2_2(s)f_1(s| T>t_0, S \leq t_0)ds+\!P(T^{(1)}\!>\!t_0, S^{(1)}\!>\!t_0) g^2_3\\
&-\!2 P(T^{(1)}>t, S^{(1)} \leq t_0) \int  g_{1}(s)f_1(s| T>t, S \leq t_0)ds\!-\!2P(T^{(1)}\!>\!t, S^{(1)}\!>\!t_0) g_{2}
\end{align*}
and $$\quad \Gbb( g) = P(T^{(0)}>t_0, S^{(0)} \leq t_0) \int   g_{1}(s) f_0(s| T>t_0, S \leq t_0) ds+P(T^{(0)}>t_0, S^{(0)}>t_0) g_{2}.$$
Taking the Frechet derivatives of the functionals, we have that for all measurable $h$ such that $ \int h^2(s) f_1(s| T>t_0, S\leq t_0) ds < \infty$,
\begin{align*}
\frac{d}{d  g_{1}} \bigg [\Lsc(g) - 2\lambda \Gbb( g)\bigg] (h)/2 &=P(T^{(1)}>t_0, S^{(1)} \leq t_0) \int  g_{1, opt}(s) h(s) f_1(s| T>t_0, S\leq t_0)ds\\
&-P(T^{(1)}>t, S^{(1)} \leq t_0) \int   h(s) f_1(s| T>t, S\leq t_0)ds \\
&- \lambda P(T^{(0)}>t_0, S^{(0)} \leq t_0) \int h(s) f_0(s| T>t_0, S\leq t_0)ds = 0.
\end{align*}
Setting $h = \delta(s)$, this implies that
\begin{align*}
g_{1, opt}(s) &= \frac{\lambda P(T^{(0)}>t_0, S^{(0)} \leq t_0) f_0(s| T>t_0, S\leq t_0)+P(T^{(1)}>t, S^{(1)} \leq t_0) f_1(s| T>t, S\leq t_0)}{P(T^{(1)}>t_0, S^{(1)} \leq t_0) f_1(s| T>t_0, S\leq t_0)}\\
&= \frac{\lambda  f_0(s, t_0, t_0)+ f_1(s, t, t_0)}{ f_1(s, t_0, t_0)},
\end{align*}
where $f_a(s, t, t_0)=P(T^{(a)}>t, S^{(a)} \leq t_0)  f_a(s| T>t, S\leq t_0)$ and $f_a(s| T>t, S\leq t_0)$ is the density of $S$ at $s$ given $(T>t, S\leq t_0, A=a)$.
And $$\frac{d}{d g_{2}} \bigg [\Lsc(g) - 2\lambda \Gbb( g)\bigg] /2= P(T^{(1)}\!>\!t_0, S^{(1)}\!>\!t_0) g_{2}-P(T^{(1)}>t, S^{(1)}>t_0) - \lambda P(T^{(0)}>t_0, S^{(0)}>t_0)  = 0,$$
which implies that
$$g_{2, opt} = \frac{ \lambda P(T^{(0)}>t_0, S^{(0)}>t_0)+P(T^{(1)}>t, S^{(1)}>t_0) }{P(T^{(1)}\!>\!t_0, S^{(1)}\!>\!t_0)}.$$
By the constraint $\Gbb( g) = \mu_0(t)$, we have
\begin{align*}
\lambda&=\bigg\{\int  \frac{f_0^2(s, t_0, t_0)}{f_1(s, t_0, t_0)}ds+\frac{P^2(T^{(0)}>t_0, S^{(0)}>t_0)}{P(T^{(1)}>t_0, S^{(1)}>t_0)}\bigg\}^{-1} \\
& \times  \bigg\{ \mu_0(t)-\int \frac{f_0(s,t_0, t_0)f_1(s, t, t_0)}{f_1(s, t_0, t_0)}ds-\frac{P(T^{(0)}>t_0, S^{(0)}>t_0)P(T^{(1)}>t, S^{(1)}>t_0)}{P(T^{(1)}>t_0, S^{(1)}>t_0)} \bigg\}.
\end{align*}

\section{}\label{appendixb}

In this section, we derive the conditions needed to guarantee the proposed PTE is between 0 and 1.
Plugging in the formula of $g_{1, opt}(s)$ and $g_{2, opt}$, we have
\begin{eqnarray*}
\Delta_{\gopt(Q_{t_0})} &=&\! \EE\{\gopt(Q^{(1)}_{t_0})\!-\!\gopt(Q^{(0)}_{t_0})\}\\
&=& \EE\{I(T>t_0)I( {S} \leq t_0) g_{1, opt}( {S})+I(T>t_0)I( {S}>t_0) g_{2, opt}|A\!=\!1\}-\mu_0(t)\\
&=&\int \frac{\lambda  f_0(s, t_0, t_0)+ f_1(s, t, t_0)}{ f_1(s, t_0, t_0)}  f_1(s, t_0, t_0) ds\\
&&+\int \frac{ \lambda P(T^{(0)}>t_0, S^{(0)}>t_0)+P(T^{(1)}>t, S^{(1)}>t_0) }{P(T^{(1)}\!>\!t_0, S^{(1)}\!>\!t_0)} P(T^{(1)}\!>\!t_0, S^{(1)}\!>\!t_0)-\mu_0(t)\\
&=& P(T^{(1)}>t)+\lambda P(T^{(0)}>t_0) -\mu_0(t). 
\end{eqnarray*}
We know that 
$ \Delta(t)=P(T^{(1)}>t)-P(T^{(0)}>t)=\mu_1(t)-\mu_0(t).$
So 
\begin{eqnarray*}
\Delta(t)-\Delta_{\gopt(Q_{t_0})}=-\lambda P(T^{(0)}>t_0).
\end{eqnarray*}
To make $\Delta(t)-\Delta_{\gopt(Q_{t_0})}>0$, we need that $\lambda<0$. 
We look into $\lambda$ further, the numerator of which is positive. The denominator of $\lambda$ is 
\begin{eqnarray*}
&&\mu_0(t)-\int \frac{f_0(s,t_0, t_0)f_1(s, t, t_0)}{f_1(s, t_0, t_0)}ds-\frac{P(T^{(0)}>t_0, S^{(0)}>t_0)P(T^{(1)}>t, S^{(1)}>t_0)}{P(T^{(1)}>t_0, S^{(1)}>t_0)}\\
&&=P(T^{(0)}>t, S^{(0)}\leq t_0)+P(T^{(0)}>t, S^{(0)}> t_0)\\
&&-\int {f_0(s,t_0, t_0)m_1(t|s, t_0, t_0)}ds-{P(T^{(0)}>t_0, S^{(0)}>t_0)M_1(t|t_0)} \\
&&=\int f_0(s,t_0, t_0)\{m_0(t|s, t_0, t_0)-m_1(t|s, t_0, t_0)\}ds+P(T^{(0)}>t_0, S^{(0)}>t_0)\{M_0(t|t_0)-M_1(t|t_0)\},
\end{eqnarray*}
where $m_a(t|s, t_0, t_0)=E[I(T>t)|S=s, T>t_0,  S\leq t_0, A=a]$ and $M_a(t|t_0)=E[I(T>t)|T>t_0, S>t_0, A=a]$.

From another angle, direct calculations show that
\begin{align*}
\Delta_{\gopt}(t_0)\!&= \!\! \EE\{\gopt(Q^{(1)})\!-\!\gopt(Q^{(0)})\}\\
&=\EE\{I(T>t_0)I( {S} \leq t_0) g_{1, opt}( {S})+I(T>t_0)I( {S}>t_0) g_{2, opt}|A\!=\!1\}\\
&-\EE\{I(T>t_0)I( {S} \leq t_0) g_{1, opt}( {S})+I(T>t_0)I( {S}>t_0) g_{2, opt}|A\!=\!0\} \\
&=  \int u f_1(u|t_0, t_0)P(T^{(1)}>t_0, S^{(1)}\leq t_0) du-\int u f_0(u|t_0, t_0)P(T^{(0)}>t_0, S^{(0)}\leq t_0)  du\\
&+g_{2,opt}\{P(T^{(1)}>t_0, S^{(1)}> t_0)-P(T^{(0)}>t_0, S^{(0)}> t_0)\}\\
&= P(T^{(1)}\!>\!t_0, S^{(1)}\!\leq \!t_0) \left[u_u\!-\!\int F_1(u|t_0, t_0) du\right] -P(T^{(0)}\!>\!t_0, S^{(0)}\!\leq\! t_0) \left[u_u \!-\!\int F_0(u|t_0, t_0) du\right]\nonumber\\
&+g_{2,opt}\{P(T^{(1)}>t_0, S^{(1)}> t_0)-P(T^{(0)}>t_0, S^{(0)}> t_0)\}\\
&= u_u\{P(T^{(1)}>t_0, S^{(1)}\leq t_0)-P(T^{(0)}>t_0, S^{(0)}\leq t_0)\}\!\\
&+\! \int\{P( U^{(1)}\!>\!u, T^{(1)}\!>\!t_0, S^{(1)} \leq t_0)\!-\!P( U^{(0)}\!>\!u, T^{(0)}\!>\!t_0, S^{(0)} \leq t_0)\}du\\ 
&+g_{2,opt}\{P(T^{(1)}>t_0, S^{(1)}> t_0)-P(T^{(0)}>t_0, S^{(0)}> t_0)\}. \label{equ2}
\end{align*}

Therefore, a set of conditions for $\Delta(t)>\Delta_g(t_0)>0$ is
\begin{eqnarray*}
(C1) && m_1(t|s, t_0, t_0)>m_0(t|s, t_0, t_0)\ \text{for all}\ s;\\
(C2) &&  M_1(t|t_0)> M_0(t|t_0);\\
(C3) &&  P( U^{(1)}\!>\!u, T^{(1)}\!>\!t_0, S^{(1)} \leq t_0)>P( U^{(0)}\!>\!u, T^{(0)}\!>\!t_0, S^{(0)} \leq t_0)\ \text{for all}\ u;\\
(C4) && P(T^{(1)}>t_0, S^{(1)}> t_0)>P(T^{(0)}>t_0, S^{(0)}> t_0),\label{conditions}
\end{eqnarray*} 
where $U=g_{1, opt}(S).$

\end{appendices}

\end{document}